\documentclass[aps,prb,preprint,showpacs,showkeys,floatfix]{revtex4}
\newcommand{\jwj}[1]{\textcolor{red}{#1}}

\usepackage{graphicx,color}
\graphicspath{{figs/}}
\begin{document}
\title{Registry Effect on the Thermal Conductivity of Few-Layer Graphene}
\author{Jin-Wu Jiang}
    \altaffiliation{Corresponding author: jwjiang5918@hotmail.com}
    \affiliation{Shanghai Institute of Applied Mathematics and Mechanics, Shanghai Key Laboratory of Mechanics in Energy Engineering, Shanghai University, Shanghai 200072, People's Republic of China}
\date{\today}
\begin{abstract}
We perform molecular dynamics simulations to study the registry effect on the thermal conductivity of few-layer graphene. The interlayer interaction is described by either the Lennard-Jones potential or the registry-dependent potential. Our calculations show that the thermal conductivity in few-layer graphene from both potentials are close to each other, i.e the registry effect is essentially not important. It is because the thermal transport in few-layer graphene is mainly limited by the interlayer breathing mode, which is insensitive to the registry.
\end{abstract}

\maketitle
\pagebreak

\section{introduction}
The two-dimensional one-atom-thick graphene sheet processes exceptional ability in heat conduction,\cite{BalandinAA2008,GhoshS2008apl} owning to ultra long life time of the flexure phonons.\cite{NikaDL2009prb,LindsayL} Its thermal conductivity is apparently higher than the three-dimensional graphite.\cite{KlemensPG} In recent years, extensive investigations have been performed to modulate the thermal conductivity in graphene through various effects.\cite{WangJ2012jpcm,JiangJW2009direction,LanJH2009,OngZY,GuoZX,YangN2009apl,JiangJW2010epl,JiangJW2010isotopic,ChenSS2012nm,YangNfold,JiangJW2011negf} We refer to Refs.~\onlinecite{Balandin2011nm,NikaDL2012jpcm,LiNB,YangN2012aipa,JiangJW2014reviewfm} for more comprehensive reviews on this topic.

As a bridge between the graphene and graphite, it is natural to imagine that the thermal conductivity in the few-layer graphene (FLG) should exhibit some dimensional crossover behavior with increasing thickness. This dimensional crossover was observed in the experiment in 2010, which shows that the thermal conductivity in FLG decreases exponentially with increasing thickness.\cite{GhoshS} The reduced thermal conductivity is attributed to the enhanced three-phonon Umklapp scattering. It results from the increased scattering channels, owning to more phonon branches in thicker FLG. This dimensional crossover phenomenon has been investigated theoretically through either the Boltzmann approach\cite{LindsayL} or the molecular dynamics (MD) simulation.\cite{SinghD,ZhangG2011nns,ZhongWR20111,ZhongWR20112,RajabpourA,CaoHY,SunT}

In all of these theoretical works, the interlayer interaction is described by the Lennard-Jones (LJ) potential $V(r)=4\epsilon ((\sigma/r)^{12}-(\sigma/r)^{6})$. The LJ potential gives good description for the cohesive behavior of graphene layers. However, it has been shown by Kolmogorov and Crespi that the LJ potential provides only small energy variation in the relative alignment of adjacent graphene layers;\cite{KolmogorovAN2000} i.e the LJ potential is insensitive to the registry. This energy variation actually should be much larger, as it is the result of the overlap between $\pi$ electrons from adjacent layers. This $\pi$-overlap is also responsible for the significant red-shift of the Raman G mode with increasing thickness of the FLG.\cite{JeonGS,JiangJW2008prb,JiangJW2008} The interlayer interaction will have direct effect on the interlayer phonon modes in the FLG. There are two groups of interlayer phonon modes in the FLG, i.e the interlayer breathing mode\cite{LuiCH} and the interlayer shear mode.\cite{TanPH} From their eigen vector morphology, it can be shown that the interlayer breathing mode is insensitive to the registry effect while the interlayer shear mode is highly sensitive to the registry effect. From above, the LJ potential can describe the interlayer breathing mode in the FLG, but it gives much lower frequency for the interlayer shear mode.

To include the registry effect, it is necessary to develop a registry-dependent (RD) potential, which contains explicit dependence on the relative shearing morphology.\cite{KolmogorovAN2000,KolmogorovAN2005} A natural question arises: is this registry effect important on the thermal conductivity of FLG? This question has not been addressed yet. We will answer it by comparatively calculating thermal conductivity of FLG with interlayer interaction described by either LJ or RD potential.

In this paper, we investigate the registry effect on the thermal conductivity of FLG, using classical MD simulations. The interlayer interaction is described by either the LJ or the RD potential. Our calculations show that the thermal conductivity from both potentials are very close to each other, which indicates that the registry effect is actually not important. It is because both potentials give similar description for the interlayer breathing mode, and the thermal transport in few-layer graphene is mainly limited by the interlayer breathing mode rather than the interlayer shear mode. We also find that the thickness dependence of the thermal conductivity in FLG is sensitive to its size and the temperature.

\section{simulation details}
\subsection{inter-atomic potential}

In our simulations, the intralayer carbon-carbon interaction is described by the Brenner potential.\cite{brennerJPCM2002} For the interlayer interaction, we employ two potentials to describe it. Similar as existing works, we first apply the LJ potential, with $\epsilon=2.5$ meV and $\sigma=0.337$~{nm}. The cut-off is 1.0~{nm}. The two parameters $\sigma$ and $\epsilon$ are fitted to experimental values for the interlayer space and the phonon dispersion along $\Gamma A$ direction in graphite.\cite{JiangJW2011mlgexpansion}

We also apply the RD potential to describe the interlayer interaction. Present calculations use the RD potential developed by Lebedeva {\it et.al}, which provides reasonable prediction of the shearing energy barrier in bilayer graphene.\cite{LebedevaIV} The interaction between two carbon atoms from different graphene layers is described by the following formula:
\begin{eqnarray}
V\left(r\right) & = & A\left(\frac{z_{0}}{r}\right)^{6}+Be^{-\alpha\left(r-z_{0}\right)}\nonumber\\
&+&C\left(1+D_{1}\rho^{2}+D_{2}\rho^{4}\right)e^{-\lambda_{1}\rho^{2}}e^{-\lambda_{2}\left(z^{2}-z_{0}^{2}\right)},
\label{eq_rdp}
\end{eqnarray}
where $r$ is the interatomic distance. $\rho = \sqrt{r^{2}-z^{2}}$ is the projection of the distance within the graphene plane, i.e the relative shearing displacement. Follow parameters are from the original work: $A=-10.510$~{meV}, $z_{0}=0.334$~{nm}, $B=11.652$~{meV}, $\alpha=41.6$~{nm$^{-1}$}, $C=35.883$~{meV}, $D_{1}=-86.232$~{nm$^{-2}$}, $D_{2}=1004.9$~{nm$^{-4}$}, $\lambda_{1}=48.703$~{nm$^{-2}$}, $\lambda_{2}=46.445$~{\AA$^{-2}$}. The interaction cut off is 1.0~{nm}.

\begin{figure}[htpb]
  \begin{center}
    \scalebox{1.0}[1.0]{\includegraphics[width=8cm]{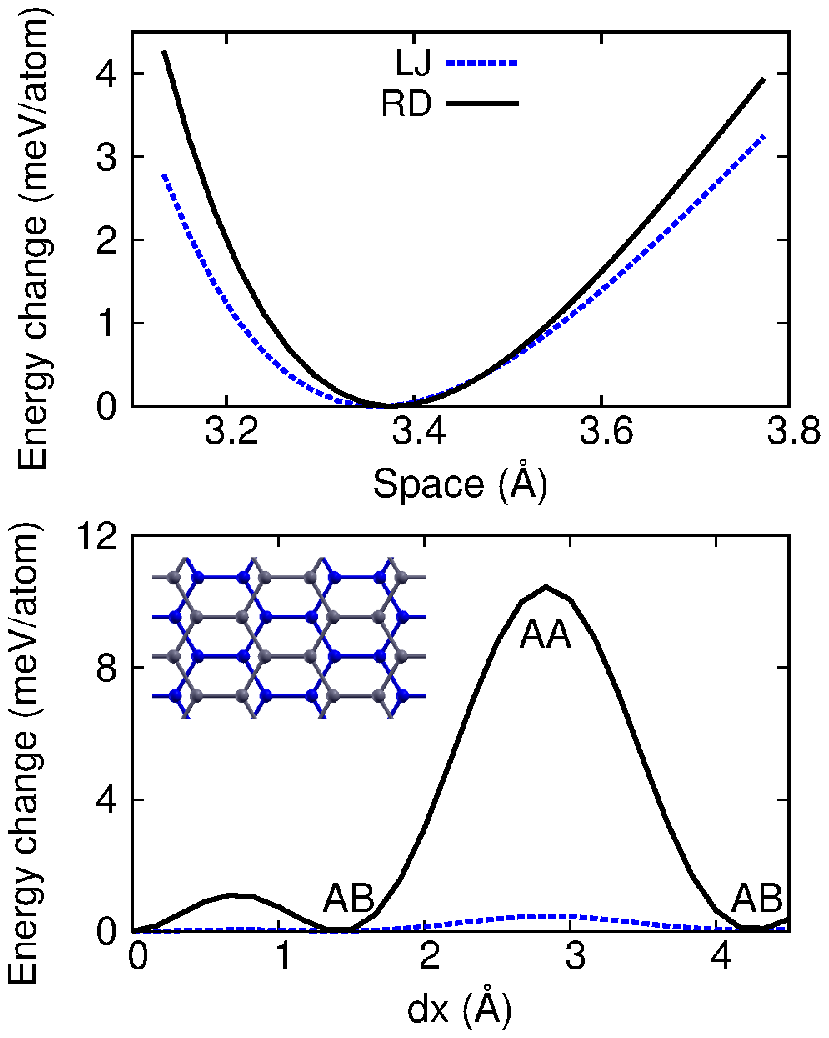}}
  \end{center}
  \caption{(Color online) The comparison between the energy per atom in bilayer graphene from LJ or RD potential. Top panel: the energy changes at different interlayer space, where the energy variation from the LJ potential is on the same order as that from the RD potential. Bottom panel: the energy also varies at different relative shift between the two graphene layers. For relative shift, the energy variation from the LJ potential is one order smaller than that from the RD potential. Inset shows the structure of the bilayer graphene at the local energy maximum point $dx\approx 0.667$~{\AA}. In both panels, the energy is relative to the global minimum energy.}
  \label{fig_rdp_lj}
\end{figure}

The LJ and RD potentials are compared in Fig.~\ref{fig_rdp_lj} for a bilayer graphene of dimension $8.5\times 1.2$~{nm}. The vertical axis is the variation of the potential energy per atom relative to the global energy minimum. From the top panel, at different interlayer space, the energy variation from the LJ potential has some difference from that of the RD potential, but they are on the same order. For the registry-related shearing of the two layers (bottom panel), the energy variation from the LJ potential is one order smaller than that from the RD potential. Inset shows the structure of the bilayer graphene at the local energy maximum point $dx\approx 0.667$~{\AA}. This comparison indicates that the LJ potential can provide a good description for those phenomena, where the registry effect is not important. However, the LJ potential gives underestimate prediction for those phenomena with important registry effect. In other words, both lJ and RD potentials give similar interlayer breathing mode in the FLG, but they provide very different prediction for the interlayer shear mode.

\begin{figure}[htpb]
  \begin{center}
    \scalebox{1.0}[1.0]{\includegraphics[width=8cm]{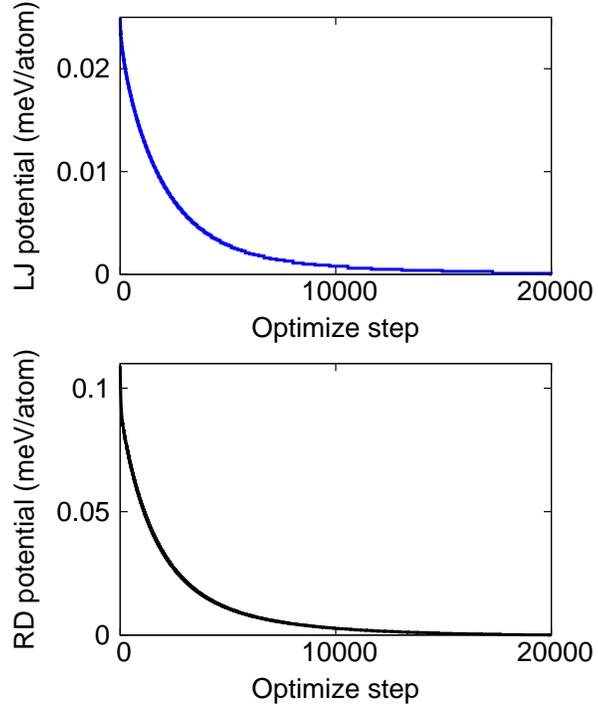}}
  \end{center}
  \caption{(Color online) The decrease of the total LJ potential or the RD potential during the structure relaxation in bilayer graphene of dimension $8.5\times 1.2$~{nm}. The structure is relaxed by the simple steepest method. The initial configuration is an AB-stacking bilayer graphene with interlayer space 0.335~{nm} and intralayer bond length 0.142~{nm}.}
  \label{fig_optimize_energy}
\end{figure}

Fig.~\ref{fig_optimize_energy} shows the evolution of the interlayer potential during the structure optimization. The simple steepest method is applied for the relaxation. The interlayer potential is described by the LJ (top panel) or the RD potential (bottom panel). It shows that the potential decreases monotonically during the whole relaxation process. This response feature manifests the consistence between the potential and the force implemented in the code.

\subsection{MD set up}

\begin{figure}[htpb]
  \begin{center}
    \scalebox{1.0}[1.0]{\includegraphics[width=8cm]{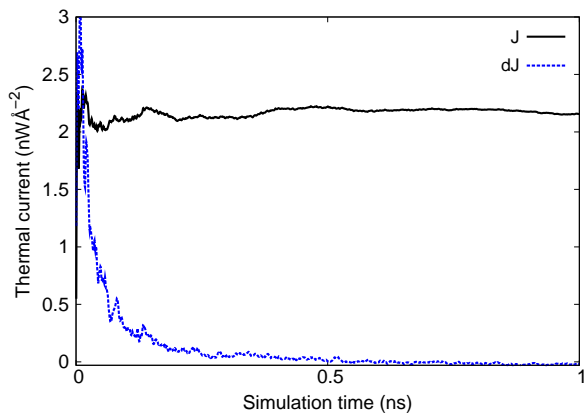}}
  \end{center}
  \caption{(Color online) Thermal current across the bilayer graphene of dimension $8.5\times 1.2$~{nm} at room temperature.}
  \label{fig_current}
\end{figure}

The heat transport is mimicked by the direct MD simulation set up.\cite{JiangJW2010isotopic} The periodic boundary condition is applied for the lateral direction. The left and right ends are fixed during the simulation. \jwj{For the two regions close to the ends, the temperature is controlled to be constant: $T_{L/R}=(1\pm \alpha)T$, with $T$ as the averaged temperature. We choose $\alpha=0.1$ in this work. The constant temperature is maintained by the No$\acute{s}$e-Hoover thermostat.\cite{Nose, Hoover} We record the energy exchange between the heat bath and the two temperature-controlled regions, say $J_{L}$ and $J_{R}$. At steady state, energy conservation requires that $J_{L}=-J_{R}$. Fig.~\ref{fig_current} shows the the thermal current flowing through the few-layer graphene, which is obtained by $J=(J_{L}-J_{R})/2$, with $dJ=J_{L}+J_{R}$ as an estimated error.}

\jwj{The Newton equations are integrated by the velocity Verlet algorithm with a time step of 1.0~{fs}. Typically, simulations are running for 10~{ns}. There are $10^7$ MD simulations steps. The system with maximum atom number in present work is the bilayer graphene with dimension $17.0\times 1.2$~nm, which has 9600 carbon atoms. The structure is thermalized to a constant temperature using the NPT (i.e. the particles number N, the pressure P and the temperature T of the system are constant) ensemble for 100~ps, so that the internal thermal pressure at finite temperature can be relaxed.}

\begin{figure}[htpb]
  \begin{center}
    \scalebox{1.0}[1.0]{\includegraphics[width=8cm]{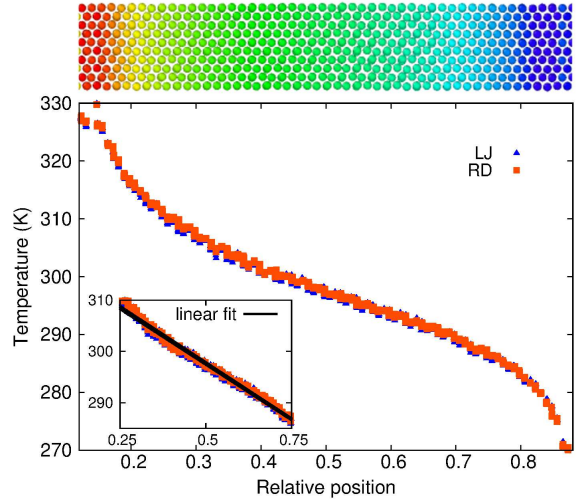}}
  \end{center}
  \caption{(Color online) The 300~K temperature profile in bilayer graphene of dimension $8.5\times 1.2$~{nm}. Temperature profiles for LJ (blue triangles) and RD (red squares) potentials are almost indistinguishable. The top inset shows the two-dimensional temperature distribution in the bilayer graphene, where atoms are colored according to their temperature (i.e averaging kinetic energy). The temperature gradient is obtained by linear fitting for the 50\% central part (left bottom inset). Both LJ and RD potentials give the same temperature gradient.}
  \label{fig_dTdx}
\end{figure}

Besides thermal current, the temperature profile is another important output from the MD simulation. Fig.~\ref{fig_dTdx} shows the temperature profile in a bilayer graphene of dimension $8.5\times 1.2$~{nm}. Temperature profiles from both LJ (blue triangles) and RD (red squares) potentials are almost indistinguishable. The right top inset shows the two-dimensional temperature distribution, where atoms are colored according to their temperature. The temperature gradient is obtained by linear fitting for the central 50\% areas (left bottom inset). The LJ and RD potentials yield almost the same temperature gradient in this case. Once the temperature gradient ($dT/dx$) is obtained, the thermal conductivity is then calculated from the Fourier law $J=-\kappa(dT/dx)$.

\jwj{It should be pointed out that the temperature profile shown in Fig.~\ref{fig_dTdx} is not purely linear. There are some nonlinear features on the left and right boundaries. Such nonlinear feature has also appeared in some previous literature.\cite{WangJS2004pre,ShiomiJ2008jjap,WangS2009jap,JiangJW2013sw} This nonlinear property is suggested to be an intrinsic property of the No$\acute{s}$e-Hoover thermostat that was used in present work.\cite{WangJS2004pre} Similar nonlinear feature has also shown up in the temperature profile with other different thermostats.\cite{JiangJW2013sw} The present work focuses on the comparison between the thermal conductivity simulated using two different inter-layer potentials (i.e. LJ an RD potentials). In both simulations, the No$\acute{s}$e-Hoover thermostat is used to set up the temperature profile, so it is expected that the nonlinear property of the No$\acute{s}$e-Hoover thermostat will have less effect on the conclusions of the present work.}

\jwj{We note that all simulations are performed above room temperature, since we are interested in the difference between the two types of interlayer potentials (LJ and RD). The quantum correction is not important for temperatures above room temperature,\cite{XuX2014nc} so we do not perform quantum corrections for the classical MD simulation.}

\section{results and discussion}

We calculate the thermal conductivity in FLG, where the interlayer interaction is described by either the LJ or the RD potential. We have shown in Fig.~\ref{fig_rdp_lj} that the energy related to the interlayer breathing movement from these two potentials is close to each other, while the energy related to the interlayer shearing movement in the FLG is quite different from LJ or RD potential. As a result, the thermal conductivity in FLG from these two potentials will be similar if the interlayer breathing mode makes dominant contribution to the thermal conductivity; i.e the LJ and RD potentials result in similar thermal conductivity if the registry effect is not important. Otherwise, quite different thermal conductivity should be obtained from the LJ and RD potentials, if the interlayer shear mode is dominates heat transport and the registry effect is important.

\begin{figure}[htpb]
  \begin{center}
    \scalebox{1.0}[1.0]{\includegraphics[width=8cm]{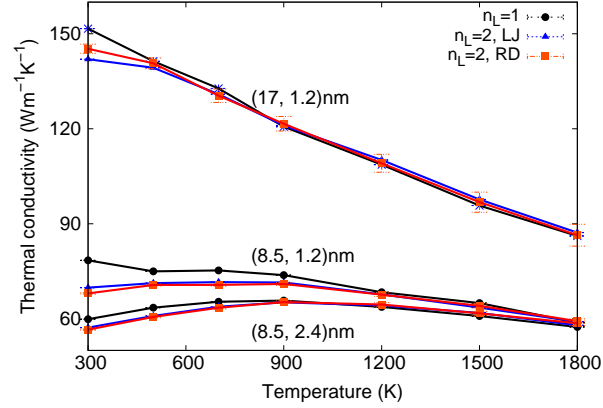}}
  \end{center}
  \caption{(Color online) The thermal conductivity versuses temperature in single-layer graphene and bilayer graphene of various dimensions. The LJ and RD potentials result in a close value of thermal conductivity in bilayer graphene.}
  \label{fig_kappa_temperature}
\end{figure}

Figure.~\ref{fig_kappa_temperature} shows the temperature dependence for the thermal conductivity from the LJ or RD potential. Three systems of different size are compared $(l_{x},l_{y})=(8.5,1.2)$~{nm}, $(2l_{x},l_{y})$, and $(l_{x},2l_{y})$. Similar as previous works,\cite{SinghD,ZhongWR20111,CaoHY} the thermal conductivity in bilayer graphene is less than that in the single layer graphene. For system with dimension $(l_{x},l_{y})$, the thermal conductivity is reduced by 11\% with layer number increasing from 1 to 2. This is much smaller than the value of about 30\% in the experiment.\cite{GhoshS} It is probably due to the size effect. The dimension of the experiment sample is on the order of micrometer. The number of the phonon-phonon scattering channel increases owning to the increased number of phonon branches in thicker FLG. This increased scattering channel directly leads to an increasing scattering rate in large samples, because there are sufficient phonon modes for the occurrence of most of the phonon-phonon scatterings. As a result, a stronger reduction in the thermal conductivity is observed in the experiment with increasing thickness. However, the system size is on the nanometer level in MD simulations, so the phonon modes are not sufficient. In this situation, the phonon-phonon scattering rate only slightly increases, although the number of the scattering channel is much greater in thicker FLG. As a result, the reduction of the thermal conductivity with increasing thickness is smaller than the experiment value. We have also calculated the thermal conductivity for a longer graphene nanoribbon of dimension $(4l_{x},l_{y})$. The thermal conductivity reduction in bilayer graphene is still around 10\%. The MD simulation becomes expensive for graphene with further increasing size; while the other approaches (eg. Boltzmann method) may be more suitable. Due to similar size effect, the thermal conductivity increases almost linearly with increasing length. The thermal conductivity in system of dimension $(l_{x},2l_{y})$ is almost double of that in the system of dimension $(l_{x},l_{y})$. For more discussions on the size effect on the thermal conductivity in nanostructures, we refer to Ref.~\onlinecite{YangN2010nt}.

\begin{figure}[htpb]
  \begin{center}
    \scalebox{1.0}[1.0]{\includegraphics[width=8cm]{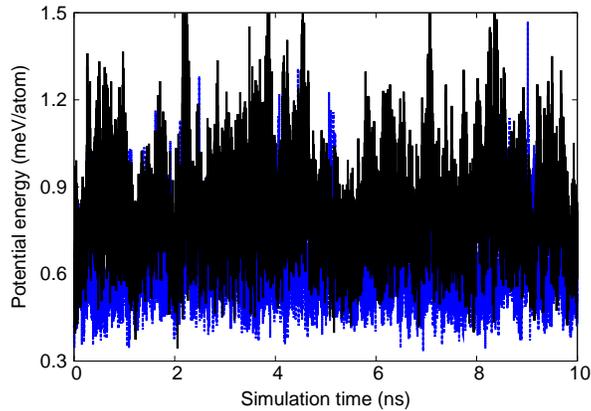}}
  \end{center}
  \caption{(Color online) The LJ (blue online) and RD (black online) potential energies during MD simulation in bilayer graphene of dimension $8.5\times 1.2$~{nm}. The energy is with reference to the value at equilibrium configuration. The RD potential energy is higher than the LJ potential, but they are on the same order.}
  \label{fig_md_energy}
\end{figure}

Let's compare the thermal conductivity from LJ and RD potentials. It is quite obvious that there is only small difference between the thermal conductivity from these two potentials. Especially for the system $(l_{x},2l_{y})$, the thermal conductivity from these two potentials are almost the same in the whole simulation temperature range. Hence, our first observation indicates that the registry effect is actually not important for the thermal conductivity in bilayer graphene. In other word, our results provide an evidence that the interlayer breathing mode plays a more important role on the thermal conductivity than the interlayer shear mode. This can be further validated by Fig.~\ref{fig_md_energy}. In the figure, the LJ and RD potential energies are compared during the whole MD simulation in bilayer graphene of dimension $8.5\times 1.2$~{nm} at 300~K. The energy is with reference to the value at equilibrium configuration. It is obvious that the RD potential energy is higher than the LJ potential, but they are on the same order. This indicates that the interlayer breathing mode is the major movement during the MD simulation. That is why the registry effect is not important. If the interlayer shear mode has major contribution, then the LJ and RD potentials shown in the figure should be very different from each other.

\begin{figure}[htpb]
  \begin{center}
    \scalebox{1.0}[1.0]{\includegraphics[width=8cm]{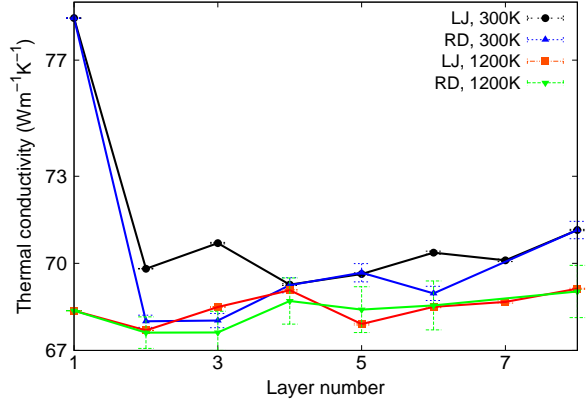}}
  \end{center}
  \caption{(Color online) The thermal conductivity v.s the number of layers in FLG of dimension $8.5\times 1.2$~{nm}. The LJ and RD potentials give close value of thermal conductivity for all FLG. Note the distinct reduction in the thermal conductivity for number increasing from 1 to 2 at 300~{K}. At high temperature (1200K), the thermal conductivity becomes insensitive to the layer number.}
  \label{fig_kappa_nL}
\end{figure}

Figure.~\ref{fig_kappa_nL} shows the thermal conductivity v.s layer number in the FLG of dimension $(l_{x},l_{y})$ at 300~K and 1200~K. First of all, similar thermal conductivity from either LJ or RD potential is obtained. It again shows that the registry effect is not important for the thermal conductivity in FLG. The thermal conductivity at 1200~K is almost the same within erors in all FLG with different layer number. At 300~K, there is a distinct reduction in the thermal conductivity when the layer number increases from 1 to 2. This is consistent with the observations in the experiment, where the reduction of the thermal conductivity mostly happens between single layer graphene and bilayer graphene.\cite{GhoshS} For layer number above 2, the thermal conductivity is almost the same, which agrees with the recent experiment by Jang et.al.\cite{JangW2013apl} In the experiment, the thinner (2-, 3-, 4-layer) graphene samples did not show any clear thickness dependence.

\section{conclusion}
\jwj{In conclusion, we have performed MD simulations to study the registry effect on the thermal conductivity of FLG. The interlayer interaction is described by either the LJ or the RD potential. Our calculations show that the thermal conductivity from both LJ and RD potentials are very close to each other, i.e the registry effect is very weak. It is because the heat transport in FLG is mainly limited by the interlayer breathing mode, which is registry insensitive. We also find that the thickness dependence of the thermal conductivity is sensitive to the size and temperature.}

\textbf{Acknowledgements} The work is supported by the Recruitment Program of Global Youth Experts of China and the start-up funding from Shanghai University.

%

\end{document}